\documentclass[a4paper]{report}
\usepackage{graphicx}
\title{Technical Note on POPI}
\author{Estel Cardellach, Serni Rib\'{o}, and Antonio Rius}
\date{\today}

\begin{document}

\begin{center}
\vspace*{4cm}
{\Huge Technical Note on POlarimetric Phase Interferometry (POPI)}\\

\vspace*{2cm}

{\Large Estel Cardellach, Serni Rib\'{o}, and Antonio Rius}\\

{\large Institut de Ci\`{e}ncies de l'Espai (IEEC-CSIC)}
\end{center}

\clearpage

\chapter{POlarimetric Phase Interferometry}

\section{Introduction}

The Global Navigation Satellite Systems (GNSS), such as the American Global Positioning System (GPS), the Russian GLONASS, or the future European Galileo, are constellations of satellites transmitting coded signals at L-band. In addition to the standard use of these signals for navigation and positioning purposes, they have also been widely used for geophysical studies (geodesy, crust and tectonics, seismics, post glacial rebound) and in radio sounding techniques along the occulting geometry (GPS-MET, Champ, Grace, SAC-C). Similarly, the potential use of these signals of opportunity as bi-static radar has arisen the interest of the scientific community because of its rich coverage (more than 50 GNSS satellites will be transmitting in a few years from now). The concept, also known as PAssive Reflectometry and Interferometric System (PARIS, \cite{mmn}), aims to use the signals after their reflection off of the Earth surface. Many experiments have been conducted to assess the possibilities and Geo-information extraction techniques of the GNSS reflected signals, mostly focused on retrieval of soil moisture, ice characterization, and preeminently, oceanographic parameters.  

The GNSS signals reflected on the sea surface have been proven sensitive to part of the sea roughness spectra (L-band roughness, for instance \cite{mebex}), and mean sea level determination (airborne altimetry at 5-cm precision level \cite{lowe}). The roughness at L-band could be relevant to air-sea interaction modelling, or for complementing and helping understanding other L-band remote sensing missions such as the ocean salinity mode of the SMOS instrument. The synergies with SMOS-like missions are currently being assessed within the CoSMOS-OS experimental campaign \cite{cosmosos}.

The purpose of this work is to present a geophysical product potentially inferable out of GNSS reflected signals: the dielectric properties of the sea surface, i.e. a combination of salinity and temperature. We present a new technique to achieve it, and study the feasibility of the concept in terms of its dynamical range and the real error budget, required system improvements, and open questions.

\section{POPI: POlarimetric Phase Interferometry}

\begin{figure} 
\includegraphics[width=12cm]{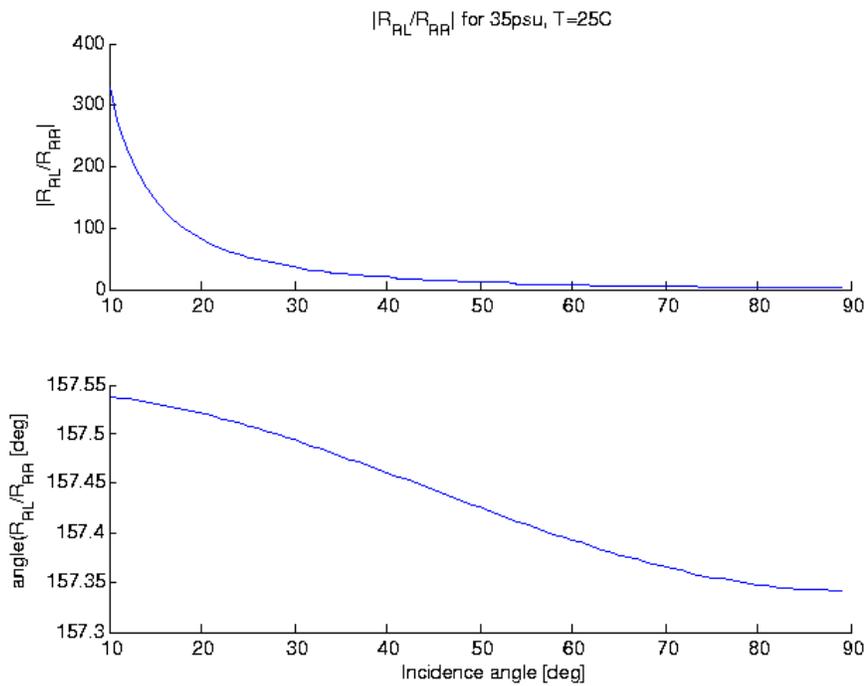}
\caption[Cross-polar and Co-polar reflection coefficients for circular polarizations]{\label{FIGfresnel} (top) Ratio between the amplitude of the Fresnel coefficients at RHCP and LHCP.  (bottom) Difference between the phases of RHCP and LHCP complex Fresnel coefficients. We have used the refined dielectric models of the sea surface at L-band in \cite{blanch}, for T=25 C and salinity at 35 psu.  }
\end{figure}

This Section presents a novel technique which aims to use the polarimetric response of the sea surface interface at L-band to infer its dielectric properties. The approach uses the {\bf phase information at two circular polarizations}. The GPS signals are emitted at Right Hand Circular Polarization (RHCP), which mostly turns into Left Hand Circular Polarization (LHCP) after the reflection off of the sea surface interface. The degree of right and left decomposition is driven by the Fresnel reflection coefficients, which in turn depend on the electrical properties of the water surface and the geometry. The comparison of the complex reflection coefficients at both circular polarizations yields a well-known amplitude ratio between the LHCP and RHCP components (Figure \ref{FIGfresnel} top), strongly dependent on the angle of incidence. However, as displayed in Figure \ref{FIGfresnel} bottom, the phase between the two polarization components sticks to a nearly constant value for all incidence angles (variation of 0.0025 degree-phase/degree-elevation).  

\begin{figure} 
\includegraphics[width=12cm]{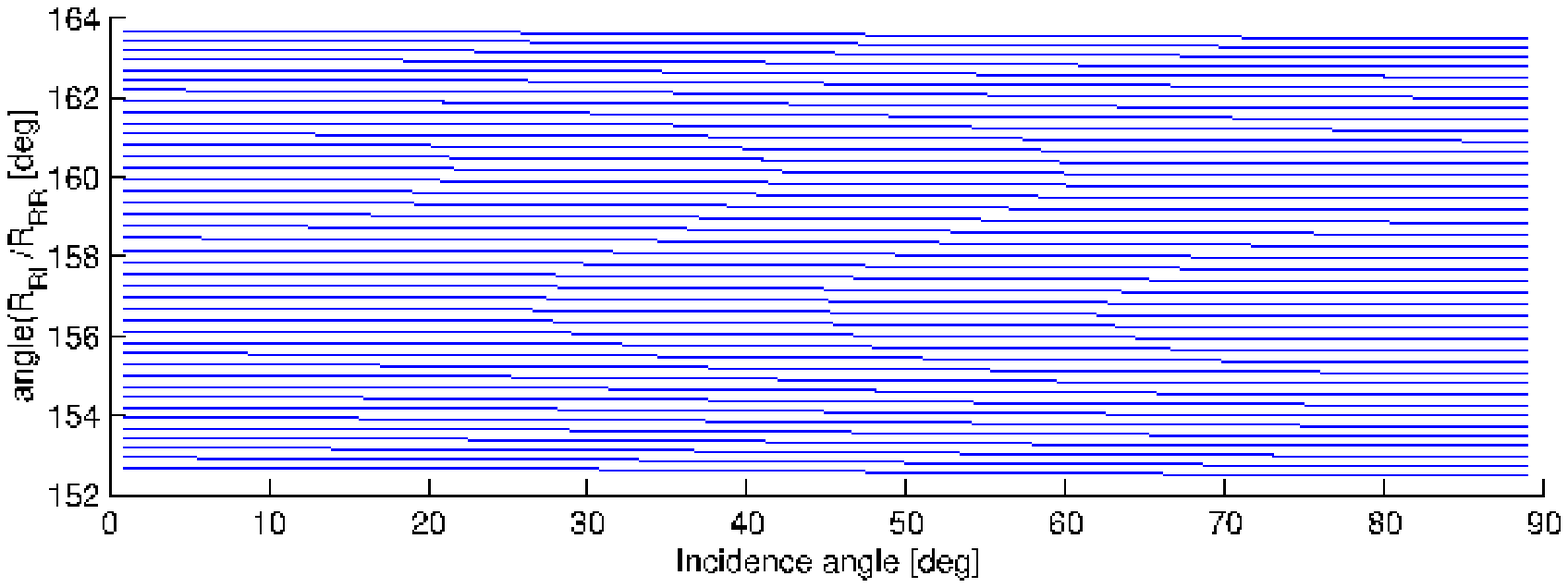}\\
\includegraphics[width=12cm]{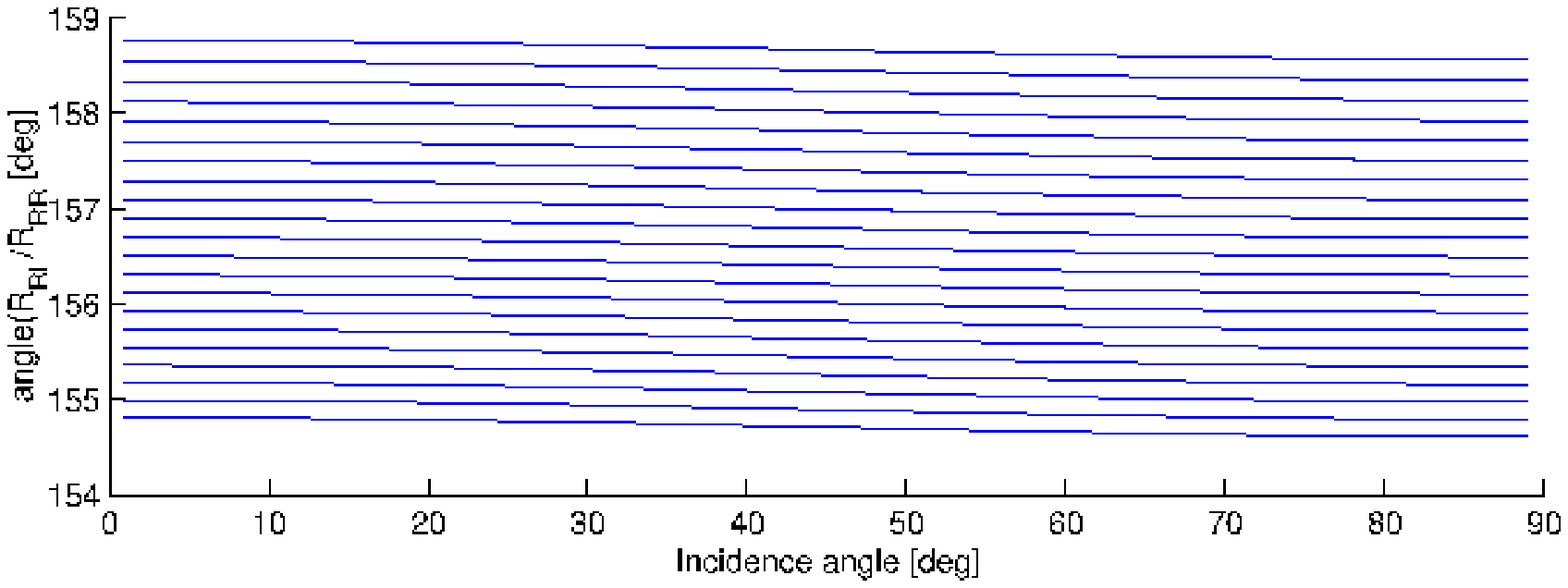}
\caption[Co- to Cross- polar relative phase according to Fresnel coefficients]{\label{FIGsaltemp} Difference between the phases of RHCP and LHCP complex Fresnel coefficients. (top) The salinity value is fixed at 35 psu, whereas the surface temperature changes from 0 to 40 C, at 1 C step. (bottom) The temperature is fixed at 25 C whilst the salinity is changed between 30 and 40 psu at 0.5 psu step.}
\end{figure}

Furthermore, the particular value of the relative RH-to-LH phase depends on the dielectric properties of the sea water (salinity and temperature), covering a dynamical range of the order of 10 degrees for sea temperature variations, and a few degrees phase variation to be sensitive to salinity (Figure \ref{FIGsaltemp}). 

The hypothesis of POPI is, therefore, that the interferometric phase between the two circular polarizations can be achieved, and linked to the water properties. The concept is easily acceptable provided that (a) the reflection essentially specular, (b) the signal is purely polarized, and (c) no other effects shift the relative RHCP-LHCP phase. Since this is obviously not realistic, point (a) is discussed under Section \ref{SECroughness}, point (b) under Section \ref{SECnopure}, and (c) is presented in Section \ref{SECothereff}. The goal is to check whether the dynamical range of the POPI phase geophysical signature ($\sim$10 degrees) is above the noise of the technique.

\subsection{Effect of the sea surface roughness}\label{SECroughness}

\begin{figure}
\includegraphics[width=14cm]{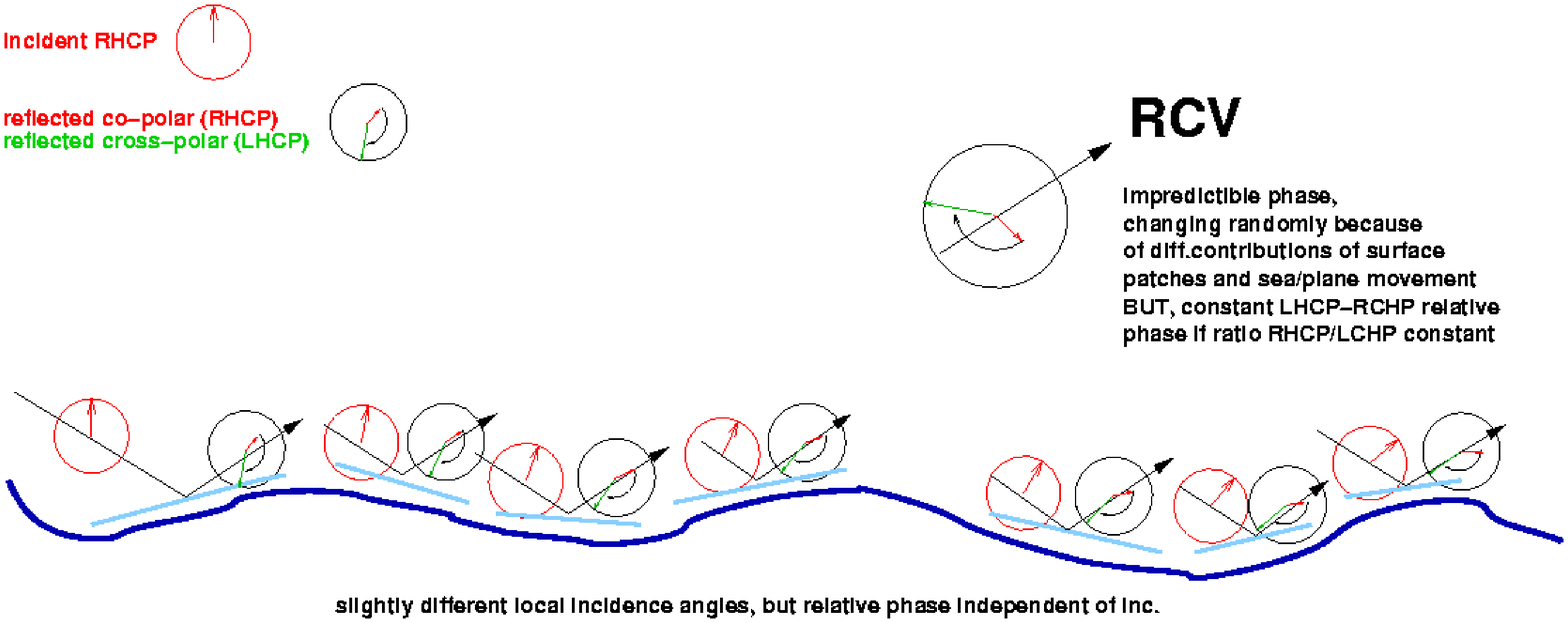}
\caption[POPI concept over rough surfaces]{\label{FIGconcept} Prevalence of POPI in spite of the sea surface roughness, here seen in the GO approximation.}
\end{figure}

The reflection of L-band signals on the sea surface is not specular solely, but it scatters energy across a wider lobe, from reflection events off of the specular. In spite of the relatively long wavelength of the L-band signals, its scattering off of the sea surface is often modelled in the Geometrical Optics (GO) limit of the Kirchoff approximation. It has been proven that GO is valid to explain a wide range of the GPS reflection behaviors (for instance, see \cite{zavor}). According to this model, the electromagnetic field received at a certain point above the surface is the sum of several field contributions, each of them coming from a specular reflection on a smooth and well oriented faced (mirror-like patch) of the surface (Figure \ref{FIGconcept}). Therefore, if we note the contribution from the $i$-facet as $E_i^{scat}=|E_i^{scat}|e^{i\Delta\phi_i^{geo-scatt}}$, the received field at $p$-polarization reads
\begin{equation}
E_r^p= \sum_i^{N-facets} |E_i^{{scat}^p}| e^{i \Delta\phi_i^{geo-scatt}}
\end{equation}
Each scattered field at each polarization $p$ relates to the incident field through the Fresnel reflection coefficients $R^p$, $E_i^{{scat}^p}=E_i^{inc} R_i^p$. The transmitted GPS signals are RHCP, meaning that the RH component of the total field after reflection comes from the co-polar coefficient ($R^{co}$), whilst the LH component is determined by the cross-polar effect ($R^{cross}$). We assume that the incident field is essentially the same on each facet, except for a phase term due to geometrical issues $E_i^{inc}=|E^{inc}|e^{i \Delta\phi_i^{geo-inc}}$,  
\begin{equation}
E_r^p= \sum_i^{N-facets} R_i^p |E^{inc}| e^{i \Delta\phi_i^{geo-scatt}} e^{i \Delta\phi_i^{geo-inc}} =  |E^{inc}|\sum_i^{N-facets} R_i^p e^{i \Delta\phi_i^{geo}}
\end{equation}
Using the Fresnel coefficients expressed in their amplitude $|R_i^p|$ and phase $\Delta\phi_i^{p}$, $R_i^p = |R_i^p| e^{i \Delta\phi_i^{p}}$, it finally results 
\begin{equation}
E_r^p= |E^{inc}| \sum_i^{N-facets} |R_i^p| e^{i \Delta\phi_i^{geo}} e^{i \Delta\phi_i^{p}}
\end{equation}

We are interested in the RH to LH relative phase, or POPI phase $\phi_{POPI}=\Delta\phi^{co}-\Delta\phi^{cross}$, which can be obtained by either complex conjugation, or its complex ratio. 

\subsubsection{Product by Complex Conjugation}

The complex conjugate product of RH and LH accounts for the correlation between them. 
\begin{eqnarray}
E_r^{RH} E_r^{{LH}^*} & =  [ |E^{inc}| \sum_i^{N-facets} |R_i^{co}| e^{i \Delta\phi_i^{geo}} e^{i \Delta\phi_i^{co}}] \nonumber  \\
& [ |E^{inc}| \sum_j^{N-facets} |R_j^{cross}| e^{-i \Delta\phi_j^{geo}} e^{-i \Delta\phi_j^{cross}}]
\end{eqnarray}
and according to Figure \ref{FIGfresnel}, $\Delta\phi_i^{co}\approx\phi_{POPI}+\Delta\phi_i^{cross}$, where $\phi_{POPI}$ take a nearly constant value for all incidence angles (therefore for all $i$-facets).
\begin{eqnarray}
E_r^{RH} E_r^{{LH}^*} &= [ |E^{inc}| \sum_i^{N-facets} |R_i^{co}| e^{i \Delta\phi_i^{geo}} e^{i \phi_{POPI}}e^{i \Delta\phi_i^{cross}} ] \nonumber \\
& [ |E^{inc}| \sum_j^{N-facets} |R_j^{cross}| e^{-i \Delta\phi_j^{geo}} e^{-i \Delta\phi_j^{cross}}] = \nonumber \\
& |E^{inc}|^2 e^{i \phi_{POPI}} [ \sum_i^{N-facets} |R_i^{co}| |R_i^{cross}| + \nonumber \\
& + \sum_{j\ne k} |R_j^{co}| |R_k^{cross}| e^{i (\Delta\phi_j^{geo}-\Delta\phi_k^{geo})} e^{i (\Delta\phi_j^{cross}-\Delta\phi_k^{cross})} ] \label{EQnocorrelation}
\end{eqnarray}
where the first term in Equation \ref{EQnocorrelation} is real, whereas the second term is negligible except for those facets geometrically correlated ($\Delta\phi_j^{geo}-\Delta\phi_k^{geo} \le 2\pi$), i.e., within the first Fresnel zone. The amplitudes $|R_j^{co}| |R_k^{cross}|$ and the phases $\Delta\phi_j^{cross}$ and $\Delta\phi_k^{cross}$ are nearly constant within the Fresnel zone, meaning that $e^{i (\Delta\phi_j^{cross}-\Delta\phi_k^{cross})}$ tends to 1, so the second terms takes the form
\begin{equation}
\sum_{j\ne k} |R_j^{co}| |R_k^{cross}| e^{i (\Delta\phi_j^{geo}-\Delta\phi_k^{geo})} e^{i (\Delta\phi_j^{cross}-\Delta\phi_k^{cross})} \sim |R_0^{co}| |R_0^{cross}| \sum_{j\ne k} e^{i \phi_{jk}^{geo}} (-\pi < \phi_{jk}^{geo} < \pi) 
\end{equation}
small compared to the first, real, term. Hence, the conjugate product becomes a phasor the phase of which is mostly driven by the POPI phase:
\begin{eqnarray}
E_r^{RH} E_r^{{LH}^*} & \sim & K  e^{i \phi_{POPI}} 
\end{eqnarray}

\subsubsection{Complex ratio}

The division reads
\begin{eqnarray}
E_r^{RH}/E_r^{LH}  & = & \frac{|E^{inc}|e^{i \phi_{POPI}} \sum_i^{N-facets} |R_i^{co}| e^{i \Delta\phi_i^{geo}} e^{i \Delta\phi_i^{cross}}}{|E^{inc}| \sum_j^{N-facets} |R_j^{cross}| e^{i \Delta\phi_j^{geo}} e^{i \Delta\phi_j^{cross}}}  \nonumber 
\end{eqnarray}
Around the nominal incidence angle, the ratio $|R_i^{co}|/|R_i^{cross}|$ takes a nearly constant value (at elevations below $\sim70^\circ$) $\forall i$  $|R_i^{co}| \sim K |R_i^{cross}|$:

\hspace*{1cm}\begin{eqnarray}
E_r^{RH}/E_r^{LH}  & = & e^{i \phi_{POPI}} K \frac{\sum_i^{N-facets} |R_i^{cross}| e^{i \Delta\phi_i^{geo}} e^{i \Delta\phi_i^{cross}}}{\sum_j^{N-facets} |R_j^{cross}| e^{i \Delta\phi_j^{geo}} e^{i \Delta\phi_j^{cross}}} = K e^{i \phi_{POPI}}
\end{eqnarray}
Note that the assumption here is stronger than those for conjugate multiplication.  If simplifications made out of this formulation are close enough to the real effects of the roughness on the scattering process, the conjugate product and/or the complex ratio would permit the extraction of $\phi_{POPI}$, which means that even if roughness adds randomness to each component of the scattered field, its relative complex conjugate product or/and ratio keeps coherent (smooth constant phase). If so, these RH+LH field combinations could be coherently integrated for long periods of time.

The anisotropies of the roughness have not been tackled in this work. Further research is required to assess its possible contribution to RH-to-LH relative phase (longitudinal wave structures imprinting certain linear polarization? i.e. a shift in the RH-to-LH phase? a few more details on this topic are under Section \ref{SECbaseline}).

\subsection{Mixed incident polarizations}\label{SECnopure}

When the incident polarization is not RH-pure, but it has a leakage of LH, and only a factor $f$ of the amplitude comes from RHCP, the incident power splits as 
\begin{eqnarray}
E_i^2 = E_i^{{RH}^2}+E_i^{{LH}^2}= f E_i^2 + (1-f) E_i^2\\
E_i^{RH}=\sqrt{f} E_i \\
E_i^{LH}= \sqrt{1-f} E_i
\end{eqnarray}
After the scattering, the mixing of polarizations reads
\begin{eqnarray}
\left[ \begin{array}{c}
E_s^{RH}\\
E_s^{LH} \end{array} \right] = \left( \begin{array}{cc} R^{co} & R^{cross} \\
                                                        R^{cross} & R^{co} \end{array} \right) \left[ \begin{array}{c} E_i^{RH} \\
	  E_i^{LH} \end{array} \right] = \left( \begin{array}{cc} R^{co} & R^{cross} \\
	                                                          R^{cross} & R^{co} \end{array} \right) \left[  \begin{array}{c} \sqrt{f} E_i \\
		     \sqrt{1-f} E_i \end{array} \right]
\end{eqnarray}
Assuming we can estimate the fraction of RH incident (from a sensitive direct RH+LH antenna, for instance)
\begin{eqnarray}
\left[ \begin{array}{c}
E_s^{RH}\\
E_s^{LH}
 \end{array} \right] = E_i \left( \begin{array}{cc} \sqrt{f} &  \sqrt{1-f}  \\
\sqrt{1-f} & \sqrt{f} 
\end{array} \right) \left[ \begin{array}{c} R^{co} \\
R^{cross} \end{array} \right]
\end{eqnarray}
the relative phase between complex $R^{co}$ and $R^{cross}$ (POPI phase, $\phi_{POPI}$) could be therefore extracted.

\subsection{Other contributions to relative phase shift}\label{SECothereff}

Besides the roughness and the mixture of polarizations, other factors might shift the RH-to-LH relative phase. For instance, the antenna pattern could affect differently each polarization, not only in terms of power, but also their phases. This would thus require absolute calibration of the involved receiving antennas, and the use of  a sensitive double polarization antenna in direct observations.

\begin{figure}[b]
\includegraphics[width=12cm]{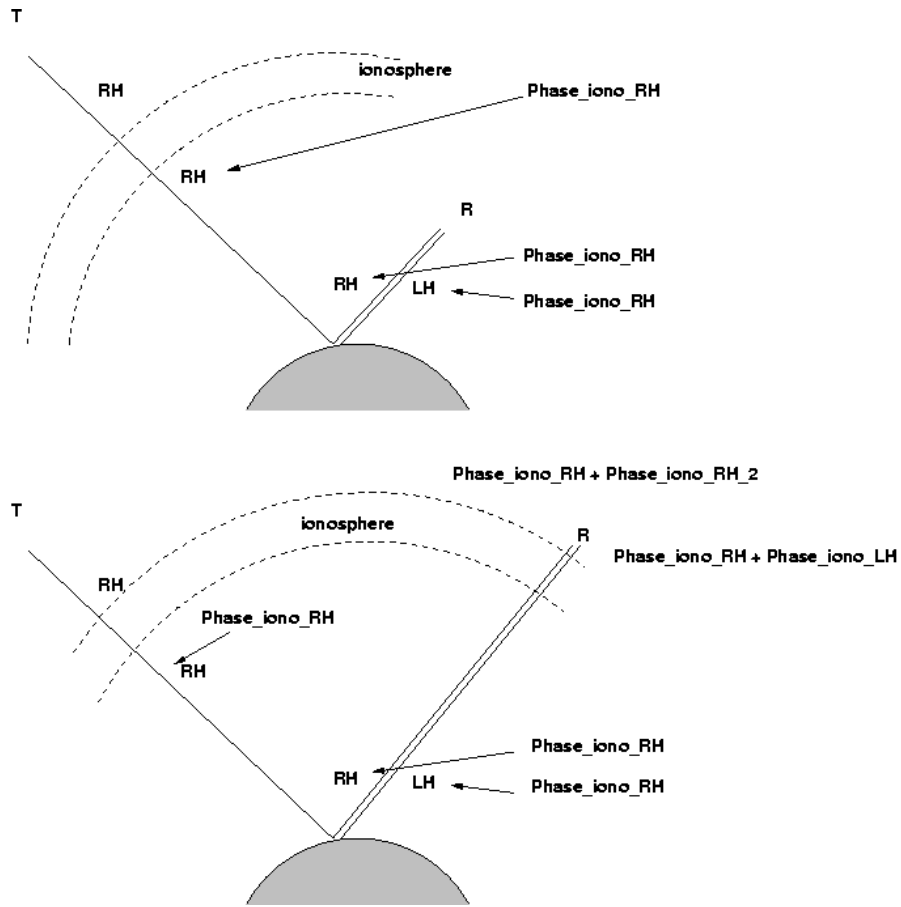}
\caption[Ionospheric effects]{\label{FIGiono} Sketch of the effect of the ionosphere when the receiver is below (top) or above (bottom) of the ionospheric layer. In the former case no corrections need to be applied to extract the POPI phase.}
\end{figure}

The ionosphere and the magnetic field provoke a Faraday rotation on the signals, inducing a different phase shift at RH and LH polarizations. Nevertheless, provided that the receiver is below the ionosphere (aircrafts, balloons, ground platforms), the signal crossing this ionized layer is the RH solely, before splitting in RH+LH at the reflection event, meaning that both LH and RH components are affected with the same ionospheric phase leap. This is illustrated in Figure \ref{FIGiono}. Note that when the receiver is above the ionosphere, the differential Fadaray rotation induced at LH and RH does need to be accounted. Possible solutions are to use multiple frequency information, or models of the electron content and the Earth magnetic field.

\chapter{POPI on real data}

Two sets of experimental campaigns have been conducted with a dedicated GPS reflections real-time hardware receiver, which provides 1-ms integrated complex waveforms, 64-lags each, from 10 channels of correlation and up to 3 simultaneous different RF inputs (antennas). The instrument, called GPS Open Loop Differential Real-Time Receiver (GOLD-RTR) can be set to acquire delay or delay-Doppler maps at L1-GPS frequency and both polarizations, with flexibility to sequentially change the settings along the experiment, with 1 Hz maximum rate of change of the configuration. More information about the GOLD-RTR can be found at \cite{nogues}.

\begin{figure}
\includegraphics[width=12cm]{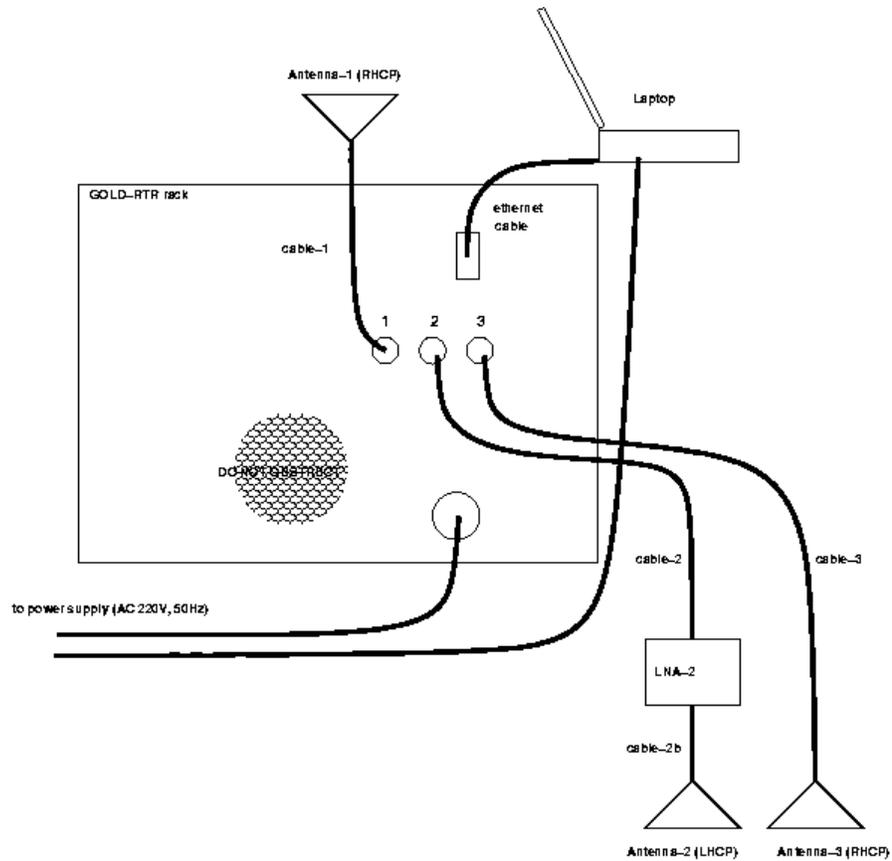}
\caption[Instrumental settings for GOLD-TEST and CoSMOS-OS campaigns]{\label{FIGsettings} Instrumental settings used to acquire the data set presented in this work. An important issue is that the down-looking antennas, used to collect GPS reflected signals, are physically located at two different spots, 12 cm a part. The geometrical projection of such a baseline into the scattering direction enters, straightforward, as a relative phase between both LH and RH signals.}
\end{figure}

The first set of campaigns is a series of 3 flights conducted in July 2005 with a CESNA CITATION I jet aircraft, flying at $\sim 10000$ m altitude and $\sim140$ m/s speed, along the Catalan Coast (North-West Mediterranean Sea), with the purpose of testing the GOLD-RTR performance (GOLD-TEST campaign, \cite{goldtest}). The later set of campaigns were conducted in the Norway Coast during April 2006: 12 flights on board a Skyvan aircraft, cruising at $\sim 3000$ m altitude and $\sim125$ m/s speed across a salinity and temperature front \cite{cosmosos}. The GOLD-RTR settings for these experiments are displayed in Figure \ref{FIGsettings}. The Skyvan was also payloaded with a infrared profiled (for sea surface temperature) and a radiometer for sea surface brightness (temperature and/or salinity). 

\begin{figure}
\includegraphics[width=12cm]{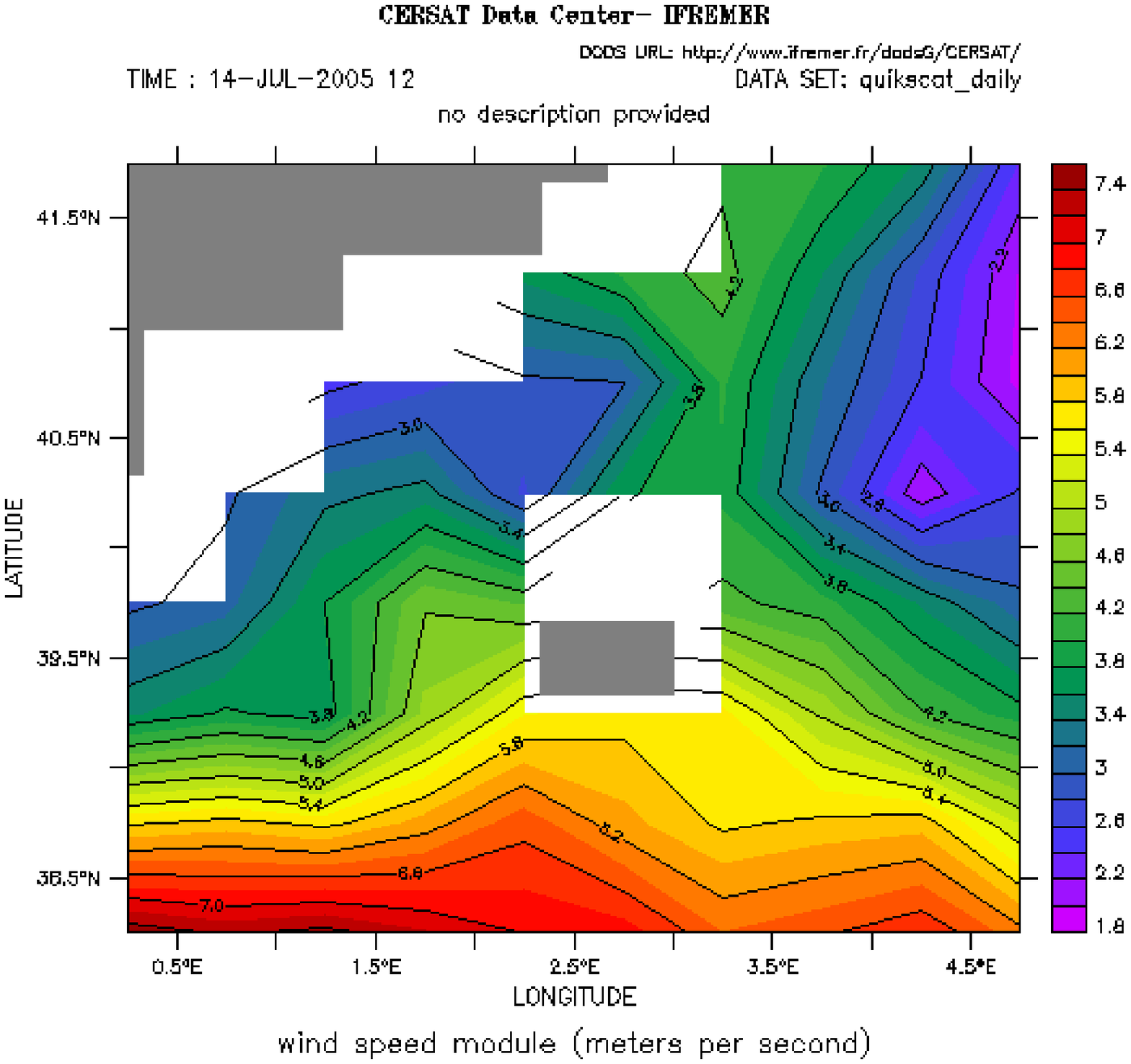}
\caption[Quikscat winds over the experimental target]{\label{FIGquikscatt} Wind conditions according to Quikscat scatterometer, orbiting above the experimental target (100 km track in front of the Catalan coast, at constant $\sim 40.7^\circ$ latitude N), i.e. mild sea roughness due to the wind. A buoy moored at latitude 41 N, longitude 1.2 E determined  $\sim$4 seconds for the waves' period. [{\it Image from IFREMER's Live Access Server}]}
\end{figure}

The current (draft) version of the Technical Note analyzes data from the 2nd flight of GOLD-TEST campaign (July 14 2006). The wind conditions over the area were mild, as showed in Figure \ref{FIGquikscatt}.

\begin{figure}
\includegraphics[width=12cm]{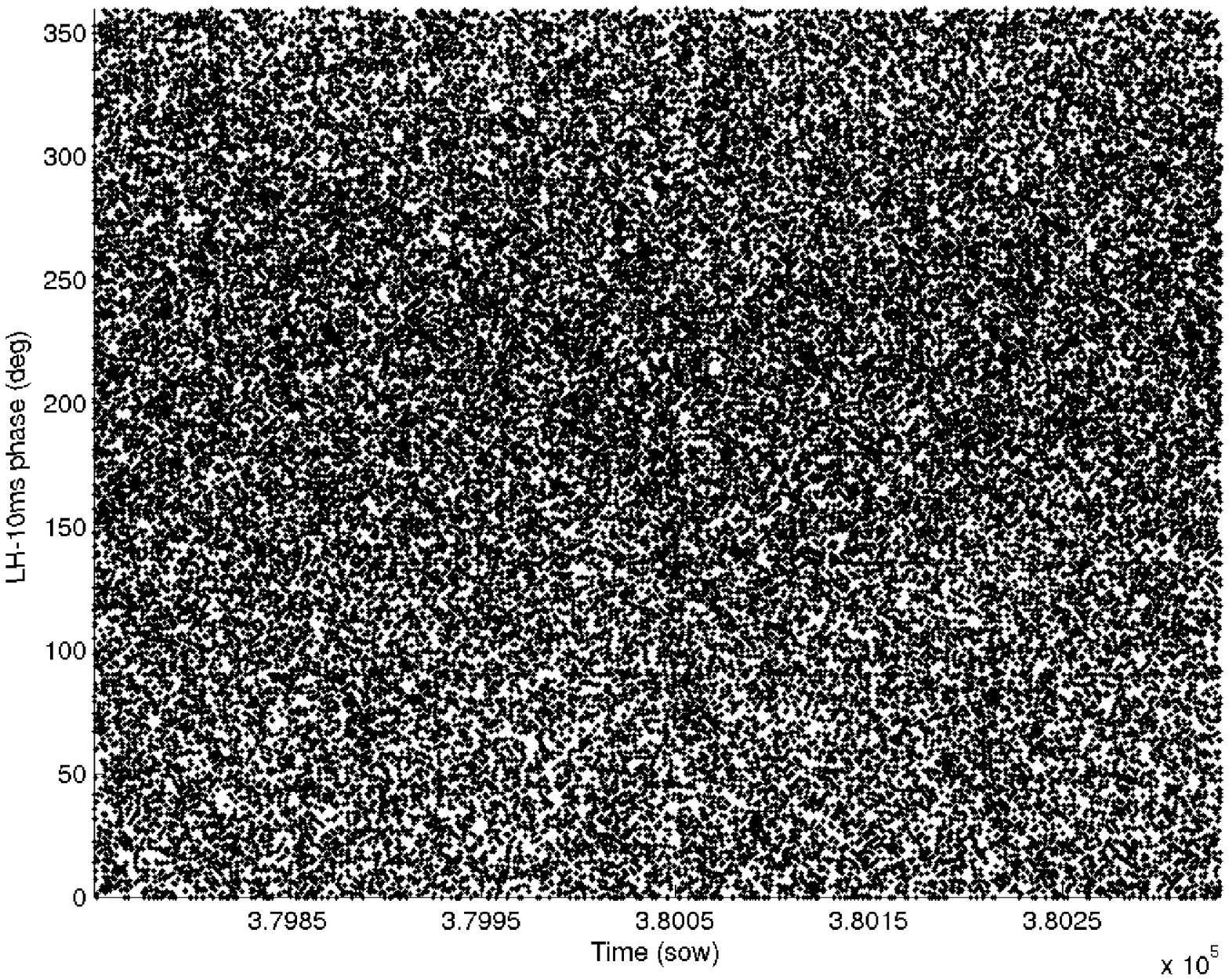}\\
\includegraphics[width=12cm]{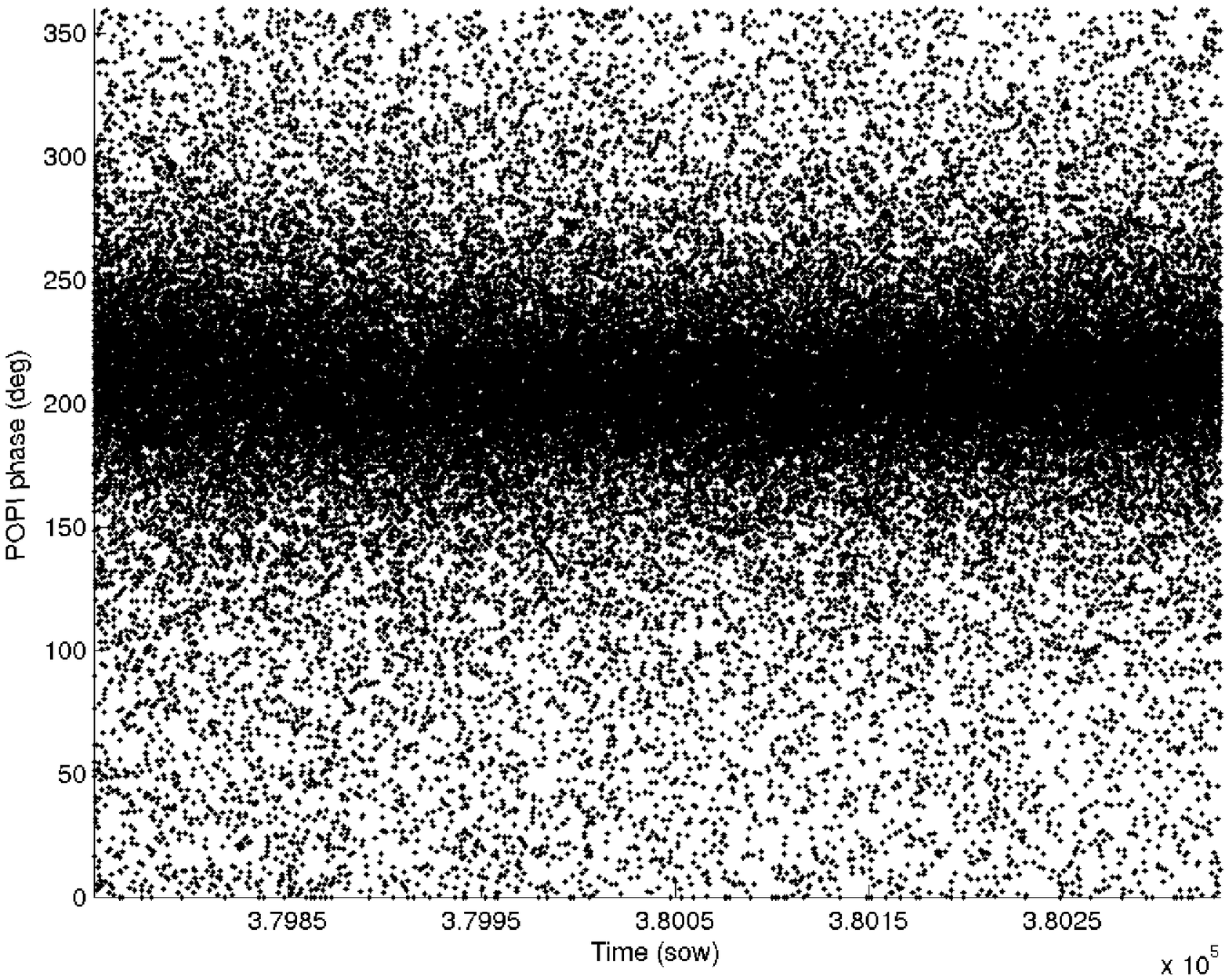}
\caption[10-ms phases for LH signal and POPI]{\label{FIGpopivsLH} (top) Time series of the 10-ms integrated LH waveforms' phase, resulting in a random sequence. (bottom) Same time series for the POPI phase, computed with the 10-ms integrated LH above and corresponding RH data ($\sim$10 minutes of PRN20).}
\end{figure}

A sample of 10 minutes data, coherently integrated at 10-ms batches, is displayed in Figure \ref{FIGpopivsLH}, showing that even if the LH signal reaches the receiver with randomly sequenced phase, the LH to RH relative phase ($\phi_{POPI}$) gets coherence.

\clearpage
\section{Conjugate Product vs Ratio}
\begin{figure}
\includegraphics[width=12cm]{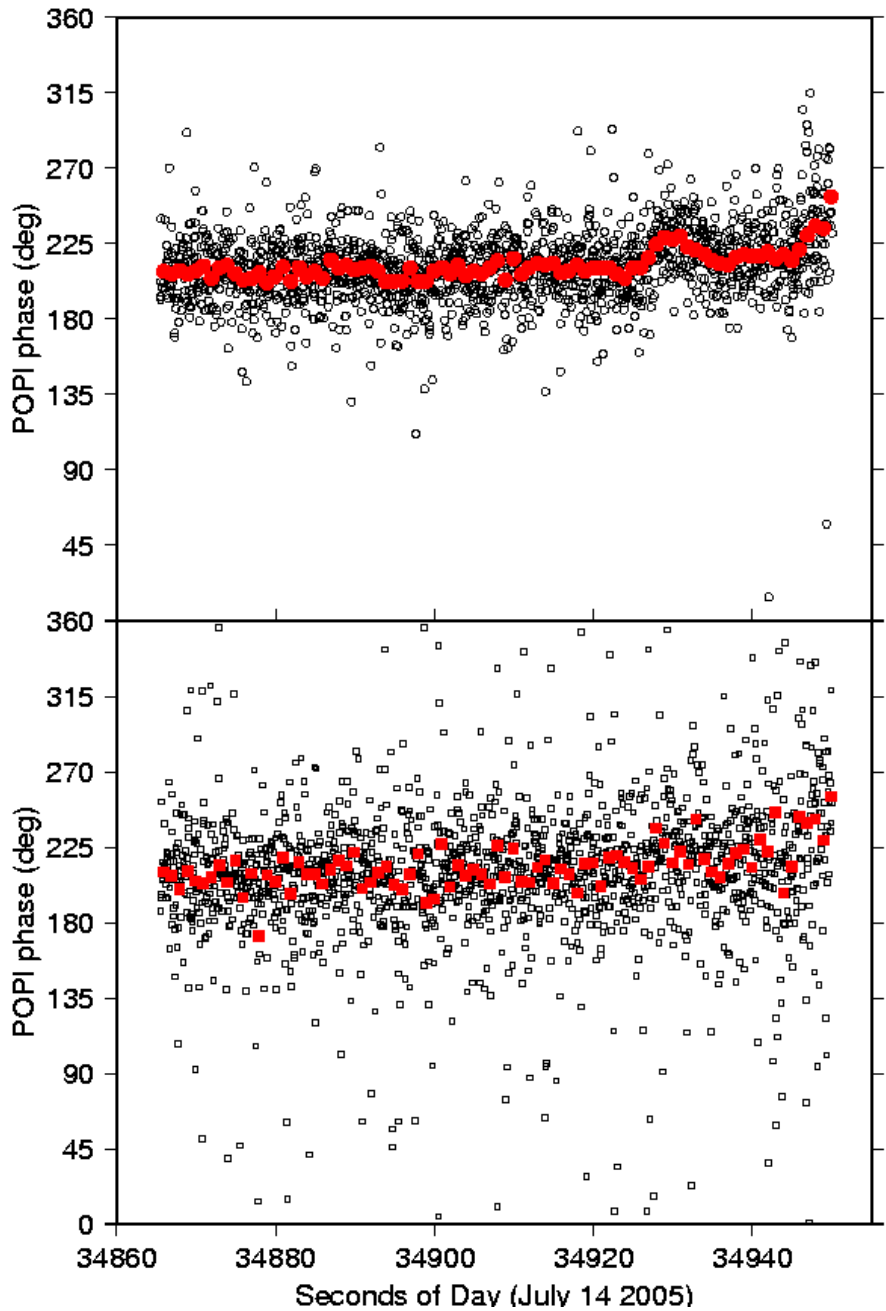}
\caption[Comparing POPI performance of product by complex conjugate vs. complex ratio]{\label{FIGconjVSratio} POPI phase computed from Conjugate Product (circles) and Ratio (squares), as complex mean of blocks of 0.05  and 1 second (empty and fill respectively), during 85 seconds of PRN20 data, which corresponds to $\sim$12 km along the surface. The performance of the Conjugate Product is more stable than the Ratio. }
\end{figure}

\begin{figure}
\includegraphics[width=12cm]{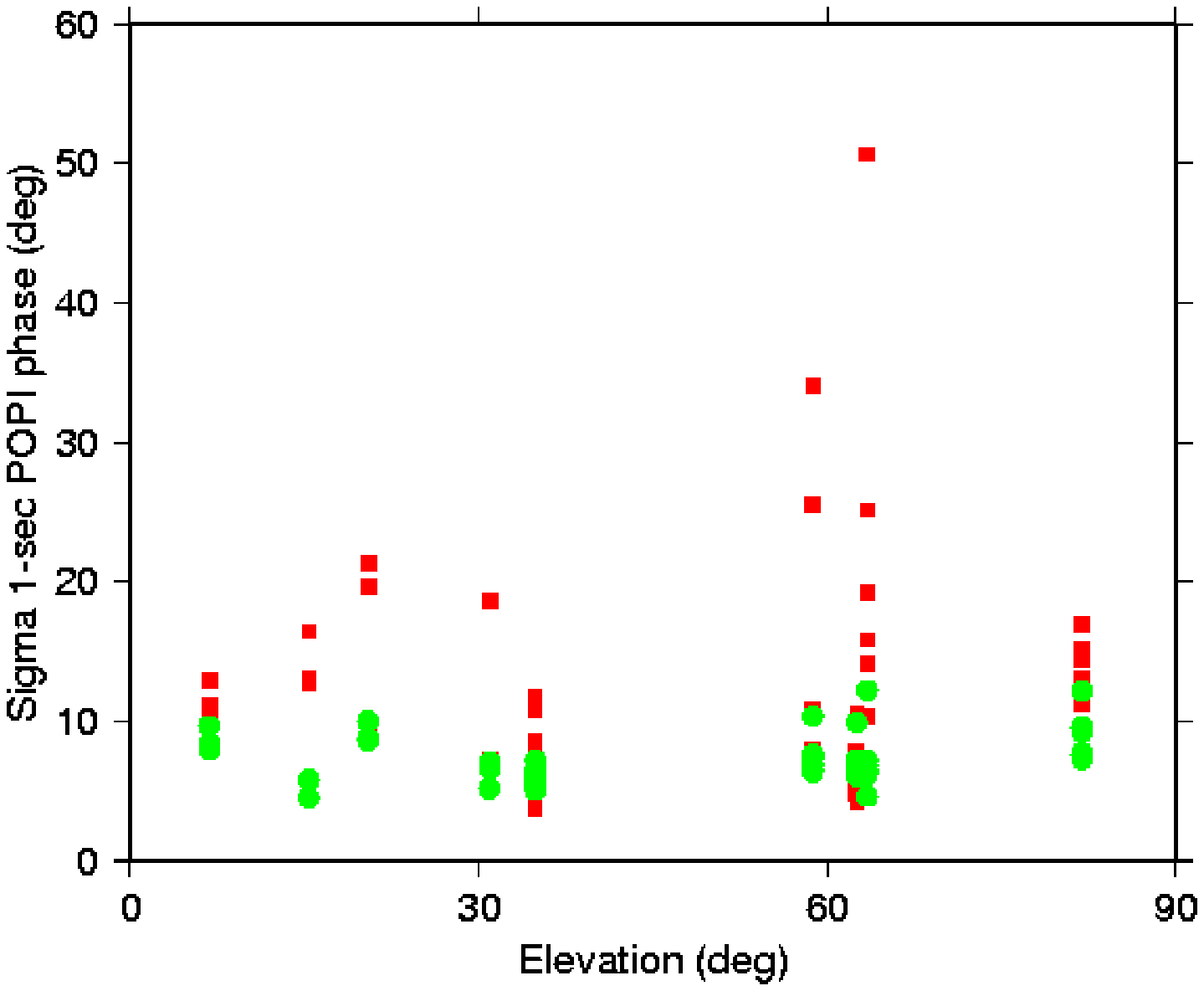}
\caption[Noise of product by complex conjugate vs. complex ratio]{\label{FIGconjVSratioELsigma} Standard deviation of the 1-sec POPI phase as function of the elevation (9 different PRNs collected during the same 10-second interval). Red squares for Ratio approach, green circles for Conjugate Product.}
\end{figure}

The POPI technique has been applied to raw data taken during the second flight of July Campaigns, using both Conjugate Product and Ratio approaches. In general for this particular case (10 km receiver altitude and mild wind) the conjugate product yields more stable and robust results than the Ratio, regardless of the elevation angle. The improvement of one approach with respect to the other increases with the accumulation time. This is illustrated in Figures \ref{FIGconjVSratio} and \ref{FIGconjVSratioELsigma}. 

\section{Coherence of POPI}

\begin{figure}
\hspace*{-3cm}\parbox{18cm}{\includegraphics[width=9cm]{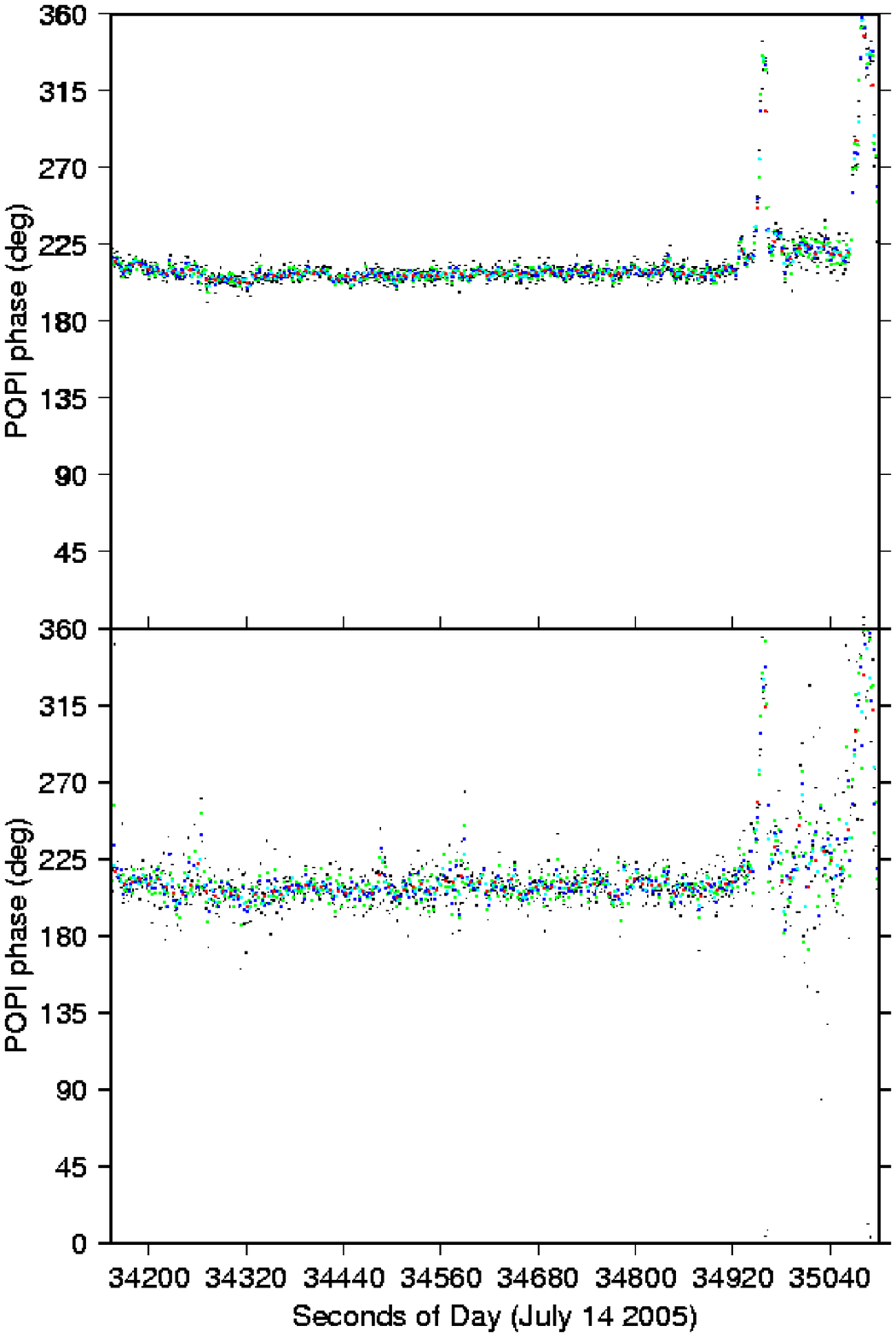}
\includegraphics[width=9cm]{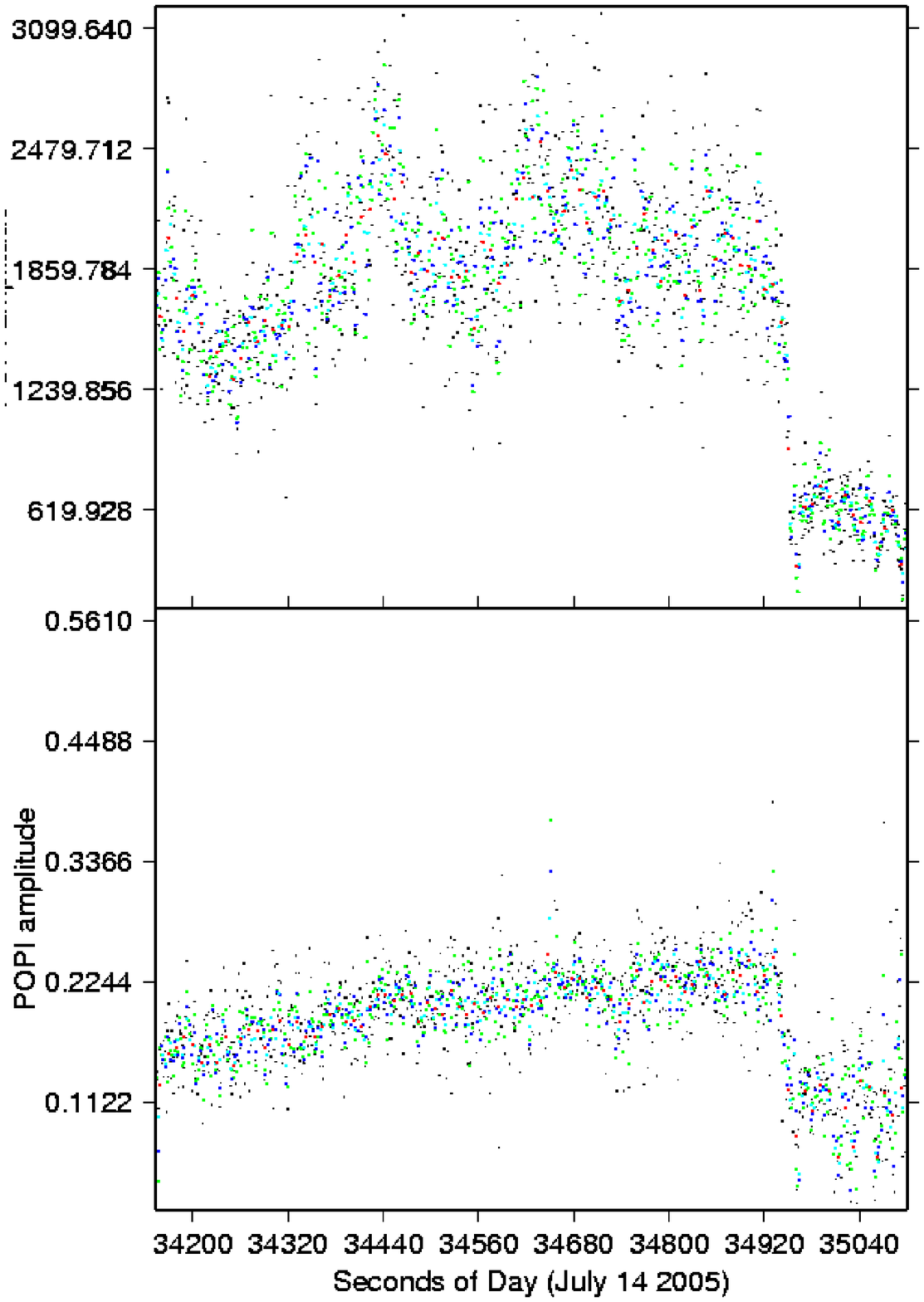}}
\caption[Example of the coherence of the POPI, when coherent integrated for 1 to 10 seconds]{\label{FIGf10prn20} POPI phase (left) and amplitude (right) as computed on $\sim$16 minutes of PRN20 as complex conjugate product (top) and complex ratio (bottom). The POPI values weakly depend on the coherent integration time (1, 2, 3, 5 and 10 seconds in black, green, blue, cyan and red respectively), a proof of its high coherence.}
\end{figure}

As described at the end of Section \ref{SECroughness}, if the assumptions under POPI were true, the POPI products would be coherent during long periods of time (long coherence between RH and LH received fields). That is, $E^{RH} E^{{LH}^*}$ and $E^{RH}/E^{{LH}}$ could be longly accumulated as complex numbers without fading effects, and its phase value would keep constant.  This is indeed true as showed in Figure \ref{FIGf10prn20}, where the POPI phase, computed from 10-ms waveforms and accumulated in the complex plane for 1 to 10 seconds (mean of the complex values) is plotted along $\sim$16 minutes, the amplitude of the POPI is also displayed, nearly equivalent regardless of the integration span.

\begin{figure}
\includegraphics[width=10cm]{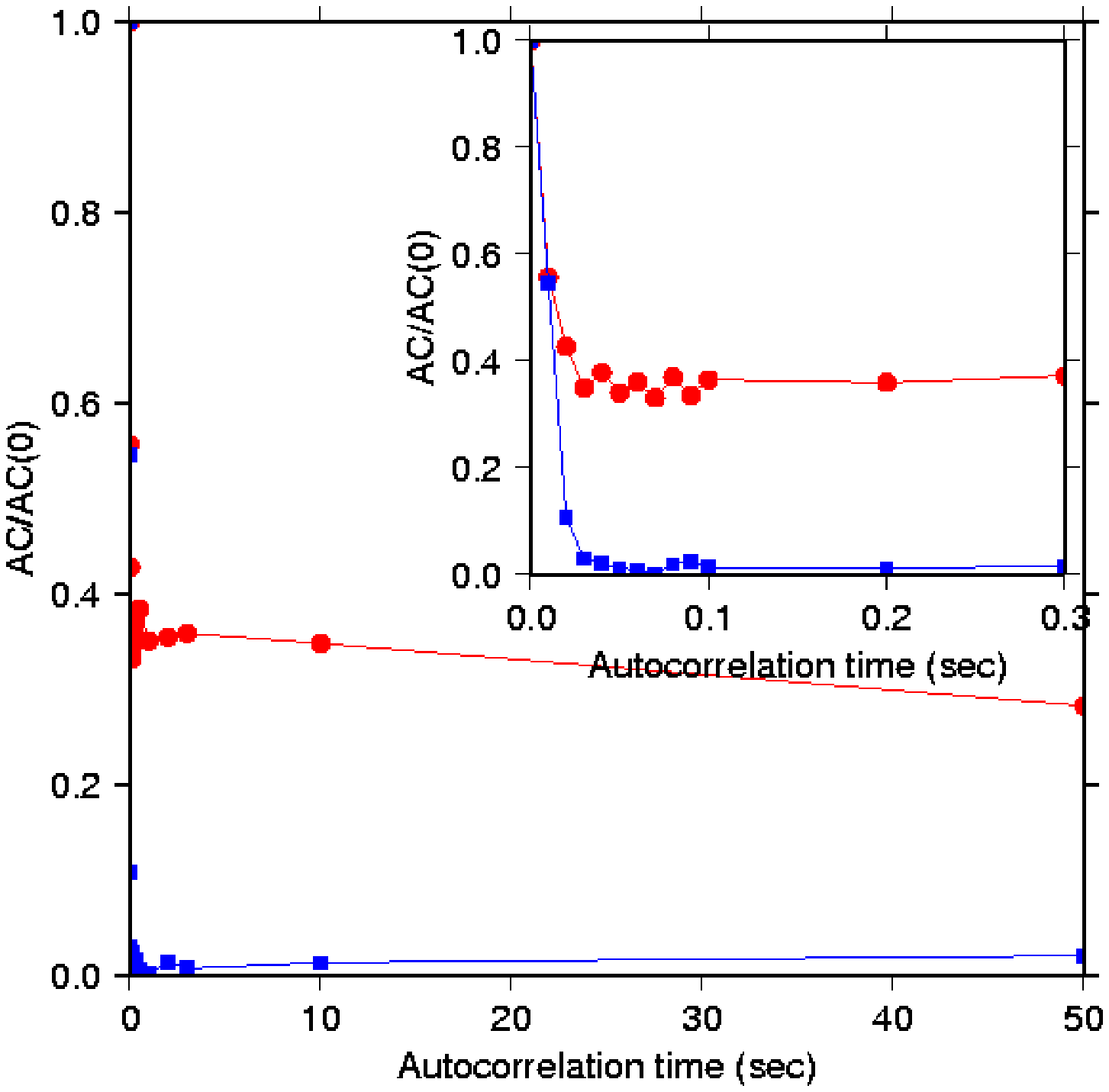}\\
\includegraphics[width=10cm]{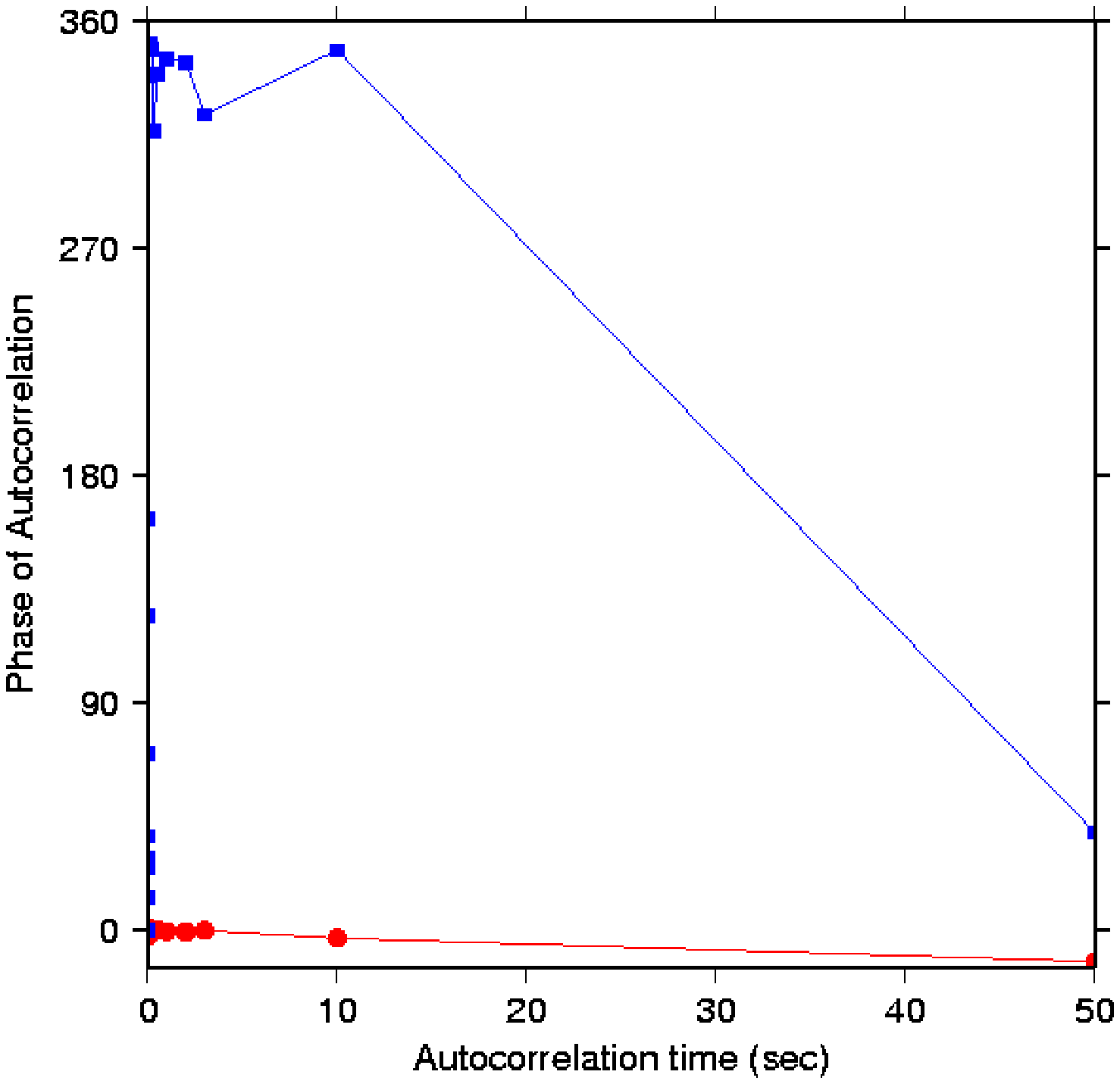}
\caption[Comparing the autocorrelation of the LH received field and POPI field]{\label{FIGsfPRN20} Equation \ref{EQsf}, normalized to $AC(\tau=0)$, computed on $T=120$ seconds of PRN20 data, blue for LH field, and red for POPI field. On top, the amplitude of $AC/AC(0)$, below, the phase of $AC$.}
\end{figure}

\begin{figure}
\includegraphics[width=12cm]{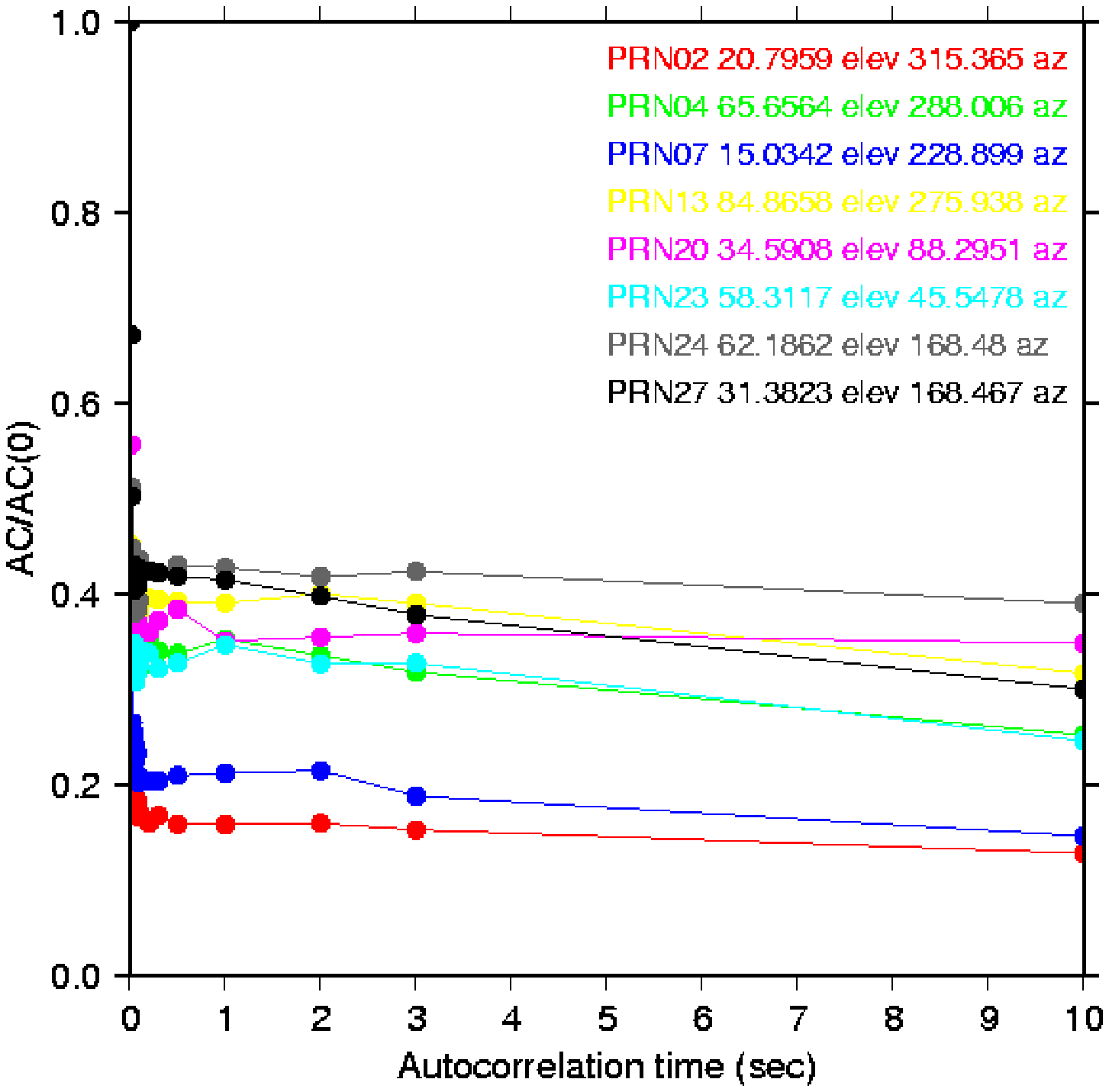}
\caption[Autocorrelation of POPI for different satellites]{\label{FIGsfALL} Autocorrelation for all satellites simultaneously acquired during 120 seconds of time.}
\end{figure}

A measure of the coherence is given by the autocorrelation function, here defined as.
\begin{equation}
AC[f(\tau)] = \frac{1}{T} \int_0^T f(t) f^*(t-\tau) dt \label{EQsf}
\end{equation}
The AC of a random field quickly drops as $\tau$ increases. The discrete evaluation of Equation \ref{EQsf} for both LH field and POPI is collected in Figure \ref{FIGsfPRN20}. The LH field loses coherence around 20-30 ms integration, whereas the POPI keeps a relatively high and constant level of coherence during a long period of time. The same figure contains the phase of the autocorrelation, which is zero for coherent series, such as POPI. The autocorrelation of the rest of the collected PRNs presents the same pattern, although the coherence level slightly changes depending on the satellite, but not on its elevation (Figure \ref{FIGsfALL}). 

\section{Precision}
\begin{figure}
\includegraphics[width=12cm]{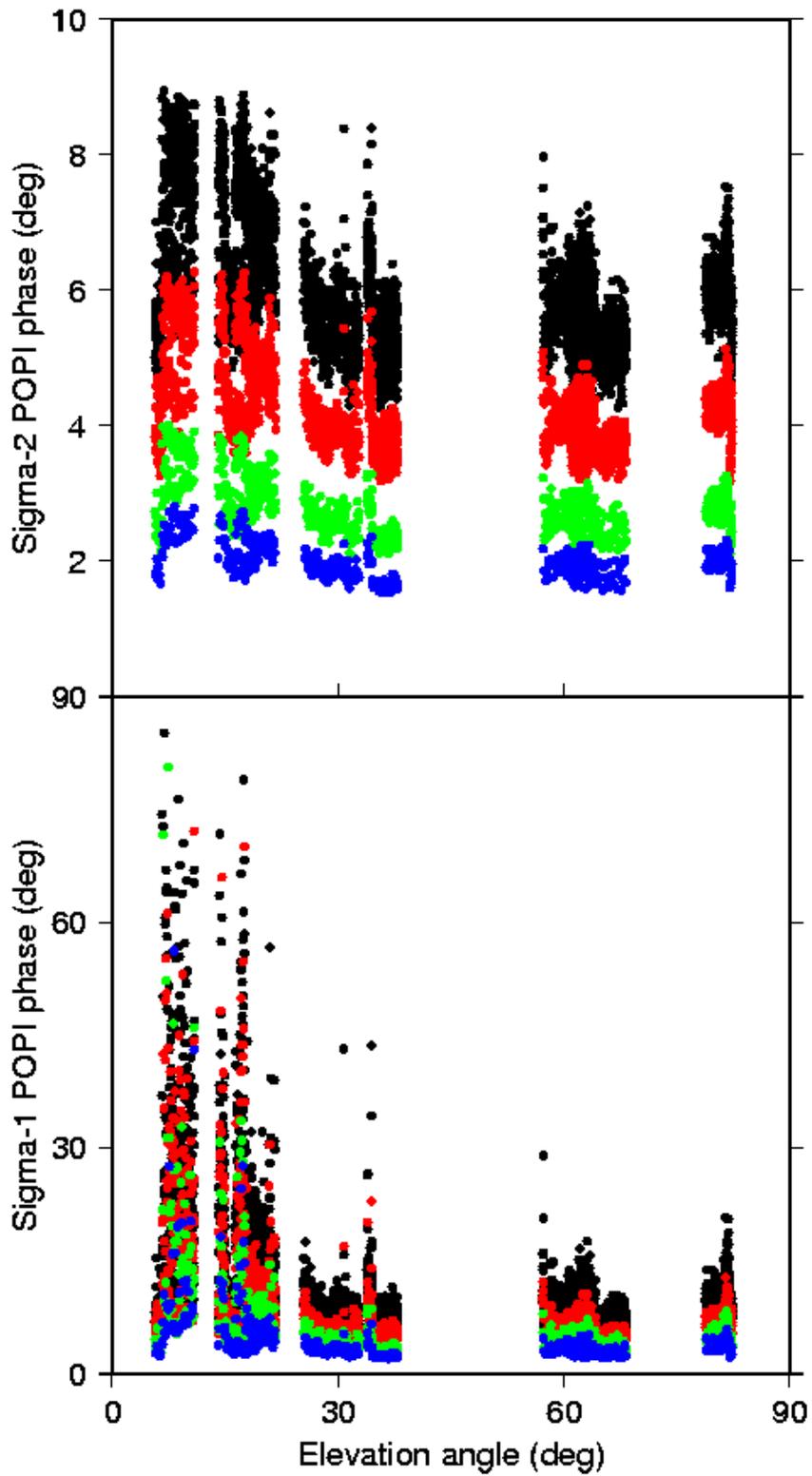}
\caption[Noise of the POPI phase as a function of the observation elevation angle]{\label{FIGsig12El} Noise of the POPI phase as a function of the observation elevation angle, computed using Equation \ref{EQsigdef1} (bottom) and Equation \ref{EQsigdef2} (top). Low angels of elevation are noisier, probably because of the worse performance of the LH polarization.}
\end{figure}

\begin{figure}
\includegraphics[width=12cm]{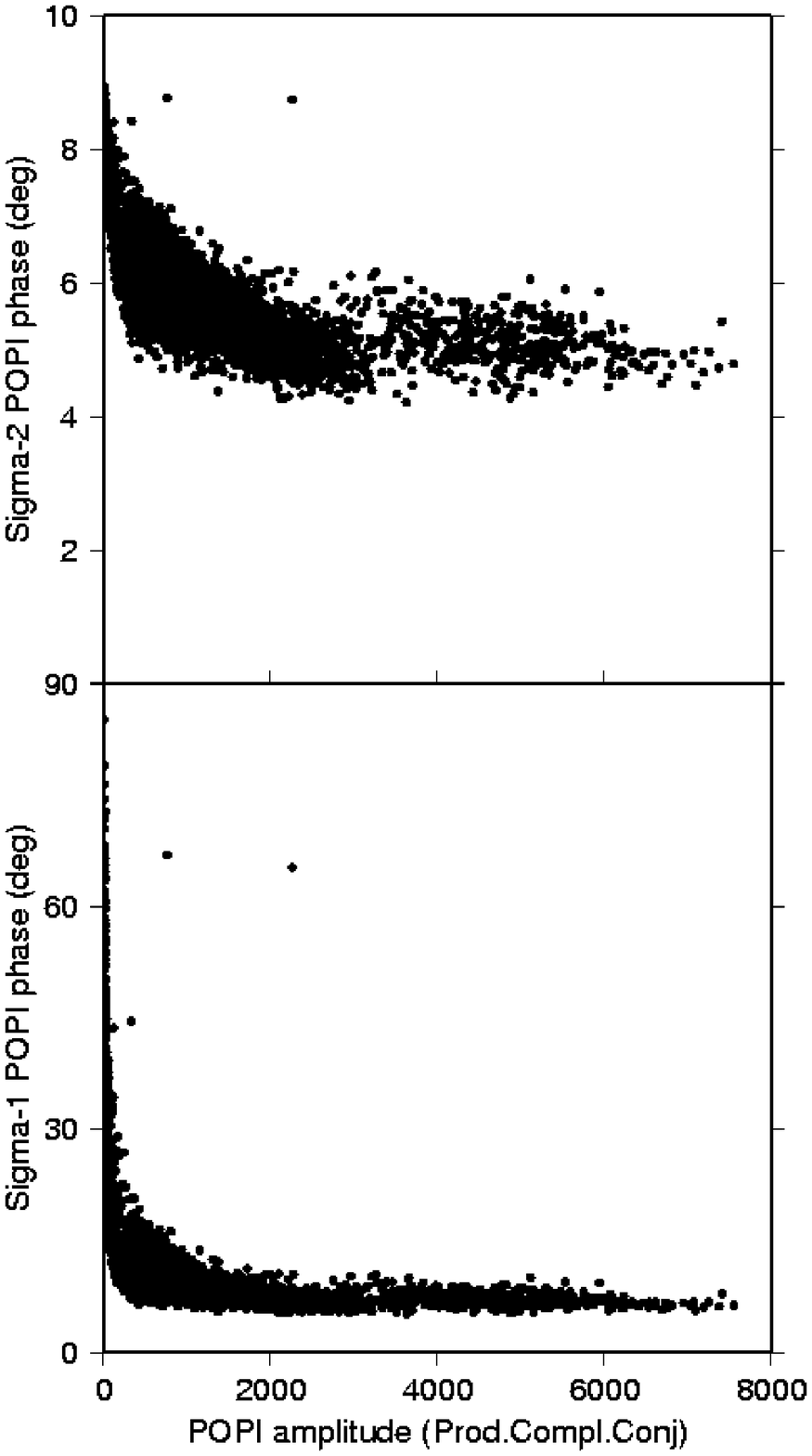}
\caption[Noise of the POPI phase as a function of the POPI amplitude]{\label{FIGsig12Ampl} Noise of the POPI phase as a function of the POPI amplitude, computed using Equation \ref{EQsigdef1} (bottom) and Equation \ref{EQsigdef2} (top), for a coherent integration time of 1 second.}
\end{figure}

The formal precision of the POPI measurements can be expressed by the standard deviation of the POPI phase, since it gives an idea of the noise of the series, its robustness, and therefore its feasibility to detect variations in the salinity/temperature (a reminder on the dynamical range: sensitivity better than 10 degrees in POPI phase is required). In general, we expect diversity of performances at different satellites, elevation angles, and integration time. Two different definitions for the $\sigma$ of the POPI phase have been used, one computed with the standard deviation of its real and imaginary parts, and the other computed as the standard deviation of the RMS-phases (understanding RMS-phases as the phases computed using the RMS spread of its real and imaginary parts). For each $i-th$ POPI sample, coherently integrated from $N_{coh}$ samples, resulting in a POPI amplitude $\rho_i$, and with real and imaginary RMS spread $rms_i^{\Re}$ and $rms_i^{\Im}$ respectively, we define
\begin{equation}
\sigma_{1_i} \equiv  atan2(\sqrt{\frac{rms_i^{\Re^2}+rms_i^{\Im^2}}{N_{coh}}},\rho_i)   \label{EQsigdef1}
\end{equation}
The second definition is the standard deviation from the derived RMS-phase dispersion:
\begin{equation}
\sigma_{2_i} \equiv \frac{1}{\sqrt{N_{coh}}} atan2(\sqrt{rms_i^{\Re^2}+rms_i^{\Im^2}},\rho_i) \label{EQsigdef2}
\end{equation}
Their means and median values are summarized in Table \ref{TABsigmas}.  As shown in Figure \ref{FIGsig12Ampl}, large sigmas  mostly correspond to cases with low POPI amplitude (low RH, LH fields, or both). These cases are concentrate at low elevation angles of observation, although not all the observations at grazing angles have high $\sigma$s (Figure \ref{FIGsig12El}). 

The conclusion from this exercise is that POPI, under the particular conditions of the actual real data set used for the analysis (10 km altitude aircraft, relatively low wind regime), is able to detect the RH to LH relative phase at a few degrees (around 3) formal precision, within the dynamical range of the global oceanic geophysical signatures. Nonetheless, this would only allow to detect variations in the sea surface temperature of the order of several degrees (6 to 15), and very weak sensitivity to salinity. The effect of the roughness and the aircraft altitude must be investigated with the data set from CoSMOS-OS campaign.

\begin{table}
\hspace*{-3cm} \begin{tabular}{|l|lll|lll|}
\hline
{\bf Int. Time (sec)} &  $\mathbf{<\sigma_1>}$  &  $\mathbf{<\sigma_1>_w}$ & $\mathbf{Median(\sigma_1)}$ &  $\mathbf{<\sigma_2>}$ &  $\mathbf{<\sigma_2>_w}$ &  $\mathbf{Median(\sigma_2)}$ \\
\hline 
 0.02 & 28.7 & 18.2 & 27.8 & 25.7 & 17.4 & 25.9 \\
 0.03 & 31.0 & 24.8 & 28.7 & 25.5 & 22.1 & 25.1 \\
 0.04 & 31.0 & 25.9 & 28.2 & 24.0 & 21.9 & 23.5 \\
 0.05 & 30.3 & 25.7 & 27.3 & 22.4 & 20.9 & 22.0 \\
 0.06 & 29.5 & 25.2 & 26.3 & 21.1 & 19.8 & 20.6 \\
 0.07 & 28.7 & 24.5 & 25.3 & 19.9 & 18.8 & 19.4 \\
 0.08 & 27.9 & 23.8 & 24.5 & 18.9 & 18.0 & 18.4 \\
 0.09 & 27.1 & 23.2 & 23.6 & 18.0 & 17.2 & 17.6 \\
 1.00 & 11.5 &  9.3 &  8.9 &  5.9 &  5.8 &  5.7 \\
 2.00 &  8.5 &  6.7 &  6.4 &  4.2 &  4.1 &  4.1 \\
 3.00 &  7.0 &  5.6 &  5.3 &  3.5 &  3.4 &  3.3 \\
 4.00 &  6.7 &  5.1 &  4.9 &  3.1 &  3.0 &  3.0 \\
 5.00 &  5.9 &  4.5 &  4.3 &  2.7 &  2.7 &  2.7 \\
 6.00 &  5.1 &  4.0 &  3.7 &  2.5 &  2.4 &  2.4 \\
 7.00 &  5.2 &  3.9 &  3.8 &  2.3 &  2.3 &  2.3 \\
 8.00 &  4.9 &  3.8 &  3.7 &  2.2 &  2.2 &  2.2 \\
 9.00 &  4.7 &  3.6 &  3.5 &  2.1 &  2.0 &  2.1 \\
10.00 &  4.5 &  3.4 &  3.4 &  2.0 &  1.9 &  2.0 \\
15.00 &  4.2 &  3.1 &  3.0 &  1.7 &  1.6 &  1.7 \\
\hline
\end{tabular}
\caption[Precision of POPI phase as function of integration time]{\label{TABsigmas} Precision of POPI phase as function of integration time. The statistical $\sigma$ has been defined as in Equations \ref{EQsigdef1} and \ref{EQsigdef2}, and the mean, weighted mean, and median have been computed from all samples ($\sim$16 minutes data set) gathered in the Table (in degrees).}
\end{table}

\clearpage
\section{Accuracy}\label{SECaccuracy}

\begin{figure}
\includegraphics[width=12cm]{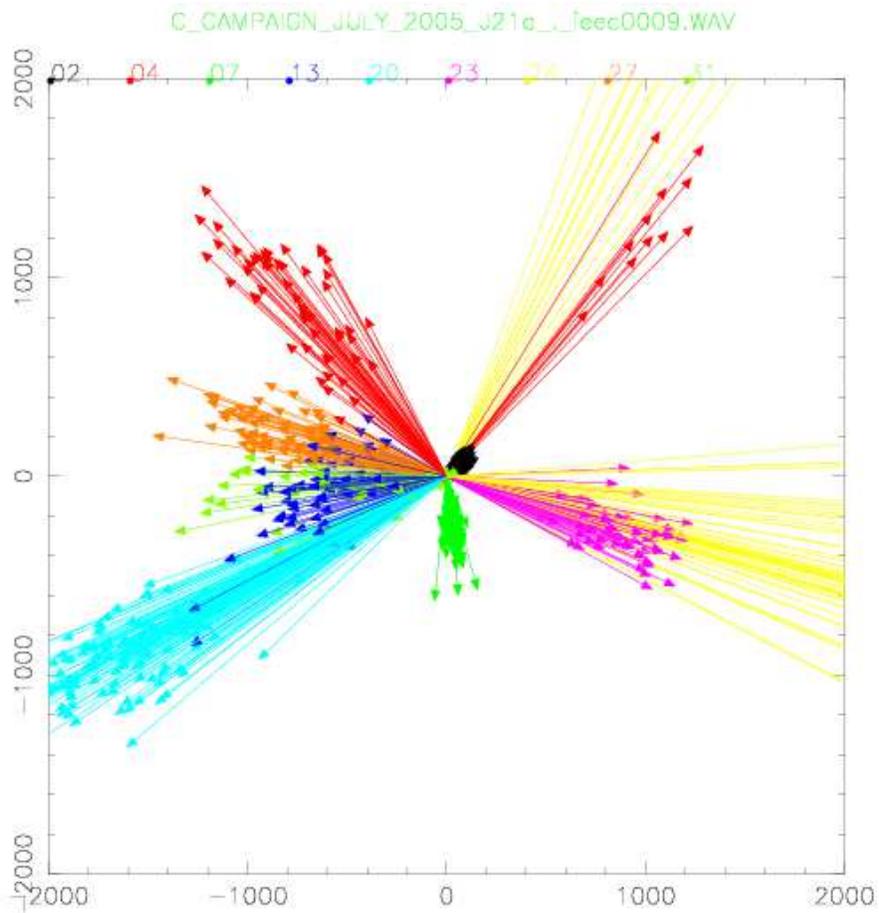}
\caption[Accuracy of POPI complex fields?]{\label{FIGvectors} This graphic contains 120 seconds of POPI complex fields, at 1 second integration, for the set of acquired satellites. The absolute phase of the POPI phasors should stick between 150 and 170 degrees, but they rather spread across the complex plane. The reasons of this inaccuracy in the absolute measurement are listed in Section \ref{SECaccuracy}.}
\end{figure}

The current instrumental implementation of the technique cannot provide absolute measurements of the polarimetric interferometric phase, required to obtain the exact values of the geophysical product. In fact, the phase POPI values, expected between 150 and 170 degrees (Figure \ref{FIGsaltemp}), take a diversity of values, as displayed in Figure \ref{FIGvectors}. The reasons are multi-fold:
\begin{itemize}
\item The RH and LH reflected signals are collected from two separate antennas, the baseline between them is therefore introducing a relative phase that depends on the geometry of the observation. This drawback could be easily solved by using a single RH+LH antenna with a minimal effective baseline between the RH and LH phase centers. More details about the baseline effect are exposed under Section \ref{SECbaseline}.
\item The antenna patterns might effect the phase differently. Currently we do not have the measurements of the actual antennas used in the experimental set up. The effects of the aircraft structure on the basis of the antenna should be investigated. Are these effects reduced when using a single bi-polarimetric antenna?
\item In case the incoming GPS signals (before scattering) are not RH-pure, which is the real incident fraction? is it significant enough to consider effects by transmitter antenna patterns? Since an absolute calibration of the transmitting antennas on-board the GPS satellites is impossible, a shortcut solution for the calibration might be the use of a single bi-polarimetric antenna for the reception of the direct links (former tests with real data using RH and LH antennas pointing to the zenith has proven it feasible \cite{ribocalibration}). This topic also arises the question about whether up to which degree the current POPI approach is indeed comparing RH and LH or a remaining LH leakage within the RH antenna. This question also requires further research.
\item Similarly, antenna effects due to near field phenomena, such as re-radiation at swapped polarization, antenna coupling, impedance mismatching...  could provoke that these particular POPI products are not really polarimetric. These problems would be solved by using a single bi-polarimetric antenna, but they hinder now the understanding of this POPI data set [{\em an indicator that the near field coupling effects are not important is to detect the matching between the baseline projection and the POPI variations, otherwise these effects would lead to another interferometric phase series, probably constant}].
\end{itemize}

\subsection{Baseline and correlators jump}\label{SECbaseline}

In our experiment, the two down-looking antennas were set at 12cm baseline (of the order of half $\lambda$). The baseline between both antennas introduce a phase given by the projection between the 3D-baseline vector ($\vec{B}$), and the unitary vector pointing from the specular reflection point towards the antenna ($\hat{u}_s$). The projection is the excess path delay between both rays, which can be expressed in degrees by means of the GPS L1 wavelength $\lambda$:
\begin{equation}
\Delta\phi_{baseline} = \vec{B} \cdot \hat{u}_s \frac{360}{\lambda} \label{EQbaselinedeg}
\end{equation}

\begin{figure}
\includegraphics[width=12cm]{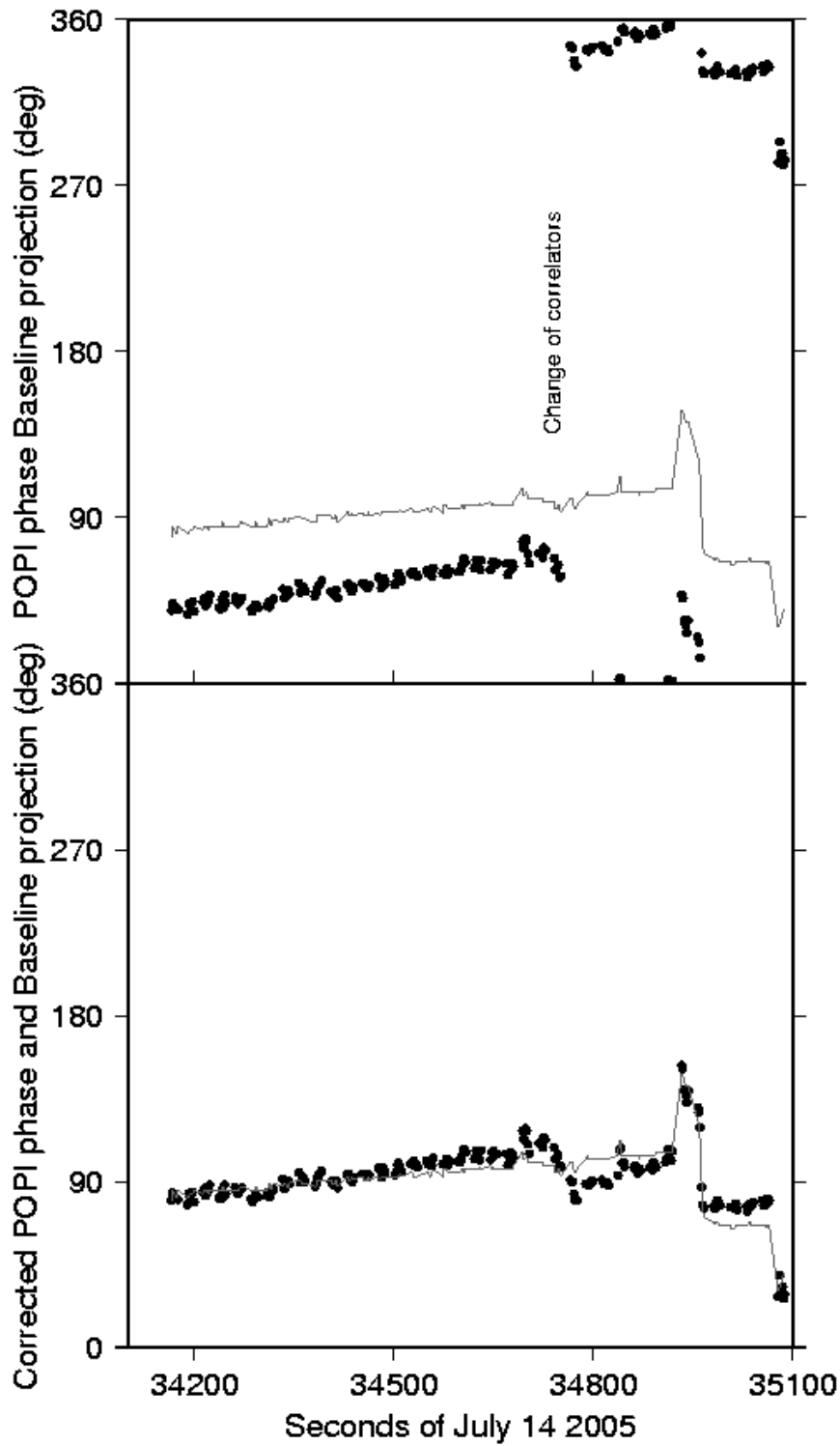}
\caption[POPI phase evolution and the corresponding baseline excess path]{\label{FIGprn24PROJ} POPI phase evolution and the corresponding baseline excess path for 1000 seconds of PRN24 data. (top) Besides the clear correlation with the baseline induced relative phase variation, the plot also shows a jump due to a change in the correlator channels. (bottom) The offsets have been readjusted to make the correlation clearer.}
\end{figure}

The aircraft was equipped with an inertial system, and the baseline vectors were precisely measured. From the combination of these chips of information, together with the GPS orbits and aircraft trajectory, we have obtained the model for $\Delta\phi_{baseline}$ along time and for each visible satellite. An example is exposed in Figure \ref{FIGprn24PROJ}, for 1000 seconds of PRN24 (July 14 2005).

Moreover, the correlation models inside the GOLD-RTR instrument use historical accumulated information for the phase. This becomes a problem when a pair of correlation channels have different history within the same run of the receiver. The current solution of this problem is to configure the data acquisition accordingly. Since it was not the case in the existing data sets, a jump in the POPI phase is detected whenever a PRN RH+LH couple switch from a pair of correlator channels to another pair.

\begin{figure}
\includegraphics[width=12cm]{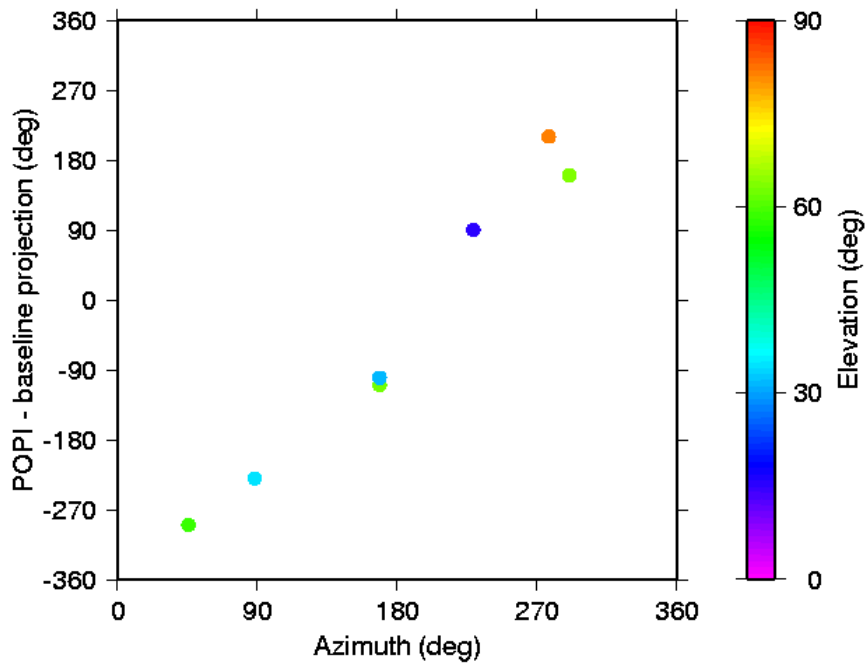}
\caption[POPI phase corrected by baseline projection, correlates with azimuth angle]{\label{FIGpopi-projAz} The first minutes of the POPI phase, before any maneuvering or swapping of correlators, have been corrected by the baseline projection. The resulting corrected POPI correlates with the azimuth angle of the scattering plane. }
\end{figure}

The first minutes of POPI angle (before any maneuvering and swapping of the correlators) have been corrected by the baseline projection. The resulting corrected POPI angles correlate with the azimuth of the scattering planes (Figure \ref{FIGpopi-projAz}). This phenomena needs further understanding, it could be related to instrumental issues (effective linear polarization of the {\it synthesized} RH+LH antenna, or/and effects of linear polarization due to longitudinal structure of the sea surface waves?).

\chapter{Summary} 

\begin{itemize}
\item A new technique has been suggested to extract dielectric properties of the sea surface by means of phase interference between the RH and LH GNSS reflected signals. The dynamical range of global sea surface temperature and salinity variations covers of the order of 10 degrees of POPI phase. Precisions below this range are required to be sensitive to this variability.
\item The technique is supported through two theoretical approaches, either the Complex Ratio of the received reflected fields or its Product by the Complex Conjugate. In both cases, the theoretical formulation requires some strong assumptions to make them work over rough surfaces. It has been proven that the product by complex conjugate yields better performance for the particular case tackled in this study (see next point).
\item The concept has been tested on real data, collected through a GPS reflections dedicated hardware receiver. The analysis presented so far comes from one flight conducted in July 2005 at $\sim$10 km altitude and mild wind conditions.
\item It has been proven that, in spite of the random behavior of each of the complex fields (LH is non-coherent after 20-30 ms complex integration), the POPI field keeps coherent for long periods of time. For instance, the autocorrelation function of the POPI scarcely drops between 20 ms and 50 seconds integration, stuck around 0.3 to 0.5 value (it depends on the satellite).
\item Such a long correlation is confirmed by the self-consistency of the POPI field, which yields nearly the same phase and amplitude values whatever complex integration period is applied.
\item The POPI phase achieves a formal precision  level of a few degrees (around 3 degrees), i.e.,  within the dynamical range of the global geophysical variations of the sea surface properties.
\item The performance under different aircraft altitudes and sea roughness states must be also assessed. The effect of roughness anisotropies must be also tackled (might longitudinal wave structures imprint certain linear polarization, a shift in the RH-to-LH relative phase?).
\item Several issues hinder the accuracy of the measurement in the current settings. Many of them would be minimized by simple improvements in the acquisition system or programming of the receiver (geometric baseline between two separate RH, LH antennas; RH+LH incident mixture; leaps due to swapping correlator channels). The others require further research, since they could constraint the final performance of the accuracy (absolute) POPI measurements, and could determine whether the results presented here correspond to POlarimetric Interferometry, or they rather correspond to the correlation with remaining LH signals in the RH-antenna output. {\bf This is the foremost open question to be solved before claiming the detection of POPI}.
\end{itemize}

\appendix

\end{document}